\ifx\undefined\documentclass
\documentstyle[cite,12pt,a4]{article}
\def\mathrm#1{{\rm #1}}
\else
\documentclass[12pt,a4paper]{article}
\usepackage{cite}
\fi

\def\Version/{2.2}
\def\Date/{December 1995}

\def\preprintno{IKDA 95/34 \\ hep-ex/9512006}

\def\hepawk/{\.{hepawk}}
\def\f77/{\.{FORTRAN-77}}
\def\hepevt/{\.{/hepevt/}}
\def\UNIBAB/{\.{UNIBAB}}
\def\UNIBABeightynine/{\.{UNIBAB89}}
\def\KRONOS/{\.{KRONOS}}
\def\KROWIG/{\.{KROWIG}}
\def\HERWIG/{\.{HERWIG}}
\def\ALIBABA/{\.{ALIBABA~2.0}}

\newenvironment{example}[2]{
    \def\examplecaption{#1}
    \def\examplelabel{#2}
    \begin{figure}
    \begin{tabbing}
    \tab\tab\tab\tab\tab\tab\tab\tab\tab\tab\kill
  }{
    \end{tabbing}
    \caption{\examplecaption}
    \label{\examplelabel}
    \end{figure}
  }

\def\.#1{{\tt#1}}         
\def\C{\`\it}             
\def\tab{\quad\=}

\def\BS/{$\backslash$}

\newcommand{\epem}{{\mathrm{e}^+\mathrm{e}^-}}
\newcommand{\Z}{{\mathrm{Z}}}
\newcommand{\dd}{{\mathrm{d}}}

\makeatletter
\def\@maketitle{\newpage \null
  \@ifundefined{preprintno}
  {\vskip 2em}              
  {                         
    \vspace*{-1\headsep}    
    \vspace*{-1\headheight}
    \begin{flushright}      
      \large \preprintno
    \end{flushright}
    \vspace*{-2em}
    \vskip \headsep         
    \vskip \headheight
    }                       
  \begin{center}            
    {\LARGE \@title \par} \vskip 1.5em {\large \lineskip .5em%
      \begin{tabular}[t]{c}\@author
      \end{tabular}\par}%
    \vskip 1em {\large \@date}
  \end{center}
  \par
  \vskip 1.5em}
\makeatother

\title{\UNIBAB/, Version \Version/: \\
       Monte Carlo Event Generation for\\
       Large Angle Bhabha Scattering\\
       at LEP and SLC Energies%
       \thanks{Supported by
               Bundesministerium f\"ur Bildung, Wissenschaft,
               Forschung und Technologie, Germany.}
}

\author{{\sc Harald Anlauf}${}^{a,b,}$%
                \thanks{e-mail: {\tt
                <anlauf@crunch.ikp.physik.th-darmstadt.de>} } \\
        {\sc Hans-Dieter Dahmen}${}^b$ \\
        {\sc Panagiotis Manakos}${}^a$ \\
        {\sc Thorsten Ohl}${}^{a,}$%
                \thanks{e-mail: {\tt
                <ohl@crunch.ikp.physik.th-darmstadt.de>}} \\
      \hfil\\
      ${}^a$ Technische Hochschule Darmstadt\\
      64289 Darmstadt, Germany\\
      \hfil\\
      ${}^b$ Universit\"at Siegen\\
      57076 Siegen, Germany\\
      \hfil\\
}

\begin{document}

\maketitle
\thispagestyle{empty}

\begin{abstract}
  This manual describes version \Version/ of the Monte Carlo event
  generator \UNIBAB/ for large angle Bhabha scattering at LEP and SLC.
  \UNIBAB/ implements higher order electromagnetic radiative corrections
  and the effects of soft photon exponentiation in a photon shower
  approach.  Weak corrections are included through the use of an
  electroweak library.
\end{abstract}


\newpage

\section*{Program Summary:}

\begin{itemize}
\item{} {\bf Title of program:} \UNIBAB/, Version \Version/ (\Date/)
\item{} {\bf Program obtainable from:} \\
        {\tt ftp://crunch.ikp.physik.th-darmstadt.de/pub/anlauf/unibab}
\item{} {\bf Programming language used:} \f77/
\item{} {\bf Computer/Operating System:} Any with a \f77/ environment
\item{} {\bf Number of program lines in distributed program, including
        test data, etc.:} $\approx$ 9000 (Including comments)
\item{} {\bf Keywords:} radiative corrections,
        large angle Bhabha scattering, multiphoton radiation
\item{} {\bf Nature of physical problem:}  Higher order leading
        logarithmic QED radiative corrections to large angle Bhabha
        scattering ($\epem\to \epem + n\gamma$)
        including weak corrections at $\Z$ energies and beyond
\item{} {\bf Method of solution:} Monte Carlo event generation
\item{} {\bf Restrictions on the complexity of the problem:} The program
        assumes dominance of the $s$-channel $\Z$ exchange contribution
        and does not include the interference between initial and final
        state radiation.  Therefore, the program is primarily designed
        to be used in the vicinity of the $\Z$ peak and only for not too
        small and not too large scattering angles ($10^\circ < \theta^*
        < 170^\circ$).  With somewhat limited accuracy, the program is
        also useful for LEP2 energies.  For cross checks, the
        contributions of the $t$-channel diagrams may be switched off,
        so that the program effectively simulates $\epem \to
        \mu^+\mu^-$.  However, the latter mode is not very efficient.
\item{} {\bf Typical running time:}  Strongly dependent on the energy
        and the chosen cuts.  The test run took approximately
         55 CPU seconds on a  DEC Alpha 3000-600 running Digital Unix,
         65 CPU seconds on an SGI Challenge running IRIX,
         80 CPU seconds on an HP 9000/735 running HP/UX,
        355 CPU seconds on an IBM RS/6000-520 running AIX, and
        485 CPU seconds on an Intel 486DX50 running Linux.
\end{itemize}


\newpage
\section{Introduction}
\label{sec:intro}

The study of electroweak physics in $\epem$-collisions close to the $\Z$
resonance has proven to be extremely fruitful \cite{AKV89,BHP95,Ols95}.
The precision measurements of the $\Z$ parameters are very powerful in
constraining the as yet unknown electroweak parameters and allow to put
stringent limits on potential physics beyond the standard model.

In the context of these precision measurements, small angle Bhabha
scattering plays a major role because all LEP/SLC experiments use it for
their luminosity measurements.  Consequently, from the beginning much
effort has gone into the construction of semi-analytical programs
\cite{BBM91b} and Monte Carlo event generators \cite{JRWW92} for Bhabha
scattering in the small angle regime.  With improved statistics,
systematic errors begin to dominate, to which the theoretical
uncertainties in the calculation of cross sections give an important, by
now even sometimes dominating, contribution.  At the level of a required
precision of the order of 0.1\%, the inclusion of higher order
electromagnetic corrections is indispensable and is taken care of in
state-of-the-art calculations and Monte Carlo programs, as reviewed in
\cite{BHP95,LEP2:BhaWG}.

In contrast, the situation for large angle Bhabha scattering is less
favorable.  In the vicinity of the $\Z$ peak, $s$-channel $\Z$ exchange
and $t$-channel photon exchange are of similar importance, thus the
calculation of radiative corrections is more involved than in the case
of fermion pair production or small angle Bhabha scattering.  For
typical experimental cuts, the radiative corrections to large angle
Bhabha scattering are numerically much larger than to small angle Bhabha
scattering, and higher order electromagnetic corrections are of much
higher importance.  Nevertheless, large angle Bhabha scattering at LEP1
and SLC in principle provides a sensitivity to the parameters of the
electroweak sector of the Standard Model comparable to the other charged
lepton final states, $\mu^+\mu^-$, $\tau^+\tau^-$.  In the energy regime
of LEP2, the contribution of $\Z$ exchange to the Bhabha cross section
is much smaller than at LEP1/SLC, hence it can be considered essentially
as a general QED test.

The state-of-the-art for large angle Bhabha scattering is reviewed in
\cite{LEP2:BhaWG}.  While semi-analytical
calculations~\cite{BBM91a,MM92,zfitter:1992,MPN+93,sabspv:1995,bhagen95}
provide reliable predictions for a predefined set of cuts, it is
nevertheless experimentally desirable to provide a comparable precision
in a full Monte Carlo event generator, which will then allow to impose
arbitrary cuts, and to simulate the response of the detector to
$\epem\to\epem$ events.

The present paper describes the implementation and status of the Monte
Carlo event generator \UNIBAB/ \Version/ that provides a level of
precision comparable to semi-analytical programs.  It is an update of
the first published version 2.0~\cite{UNIBAB2.0}, and has been compared
with other available Monte Carlo event generators ({\tt
BHAGENE3}~\cite{bhagene3:1995} and {\tt BHWIDE}~\cite{bhwide:1995}) as
well as available semi-analytical calculations~\cite{LEP2:BhaWG} both
for LEP1 and LEP2 energies.

The distinguishing feature of \UNIBAB/ from fixed order Monte Carlo
event generators is that it handles the leading logarithmic
contributions
\begin{equation}
\label{eq:leading-logs}
  \frac{\alpha}{\pi} \log \left(\frac{s}{m_{\mathrm{e}}^2}\right)
  \approx 6\%   \qquad  \mbox{(at LEP/SLC energies)}
\end{equation}
to the electromagnetic radiative corrections to all orders in a Monte
Carlo parton shower algorithm, which includes the very important
exponentiation of the soft photon contributions automatically, as well
as the contributions from multiple emission of hard collinear photons.
In addition, \UNIBAB/ contains the full kinematics from multiphoton
emission, since the radiated photons are generated explicitly.

The first version of the Monte Carlo generator, \UNIBABeightynine/
\cite{DMMO89,DMMO91}, was a ``QED dresser'' in the sense of
\cite{AKV89}, i.e.\ it contained only the QED corrections to an
effective Born cross section.  It also suffered from several serious
limitations (see appendix~\ref{sec:History}).  In version 2.0 of
\UNIBAB/, most of these limitations were overcome using the experiences
with the Monte Carlo event generators \KRONOS/ \cite{ADM+92a} and
\KROWIG/ \cite{Ohl92b} for deep inelastic scattering at HERA.  Weak
corrections are included through the use of an electroweak library.

This write-up is organized as follows: In section \ref{sec:bhabha} we
outline the physics underlying the algorithms implemented in \UNIBAB/.
The actual implementation is described in section
\ref{sec:implementation}.  The parameters controlling the execution of
\UNIBAB/ are discussed in detail in section \ref{sec:parameters}, and
the \f77/ interface is presented in section \ref{sec:f77}.  Section
\ref{sec:concl} contains our conclusions.  Distribution rules, the
revision history, a listing of all external symbols and an example can
be found in the appendices.


\section{Bhabha Scattering at High Energies}
\label{sec:bhabha}

In the structure function formalism \cite{BBM91a,KF85,AM86,BBN89} the
factorized expression for the differential cross section for the
process
\begin{equation}
  \mathrm{e}^+(p_+) \mathrm{e}^-(p_-)
    \to  \mathrm{e}^+(p'_+) \mathrm{e}^-(p'_-)
         \left[ \gamma(k_1) \cdots \right]
  \label{eq:momenta}
\end{equation}
reads
\begin{eqnarray}
  \label{eq:factorization}
  \frac{\dd^4\sigma}{\dd t_+ \dd t_- \dd u_+ \dd u_-}
    & = & \int\limits_0^1 \dd x_+ \dd x_- \dd y_+ \dd y_-
          \Gamma(x_+,Q^2) \Gamma(x_-,Q^2)
          D(y_+,Q^2) D(y_-,Q^2) \nonumber \\
    &   &\times\;\frac{\partial (\hat t_+, \hat t_-, \hat u_+, \hat u_-)}
                      {\partial (t_+, t_-, u_+, u_-)}
         \;      \frac{\dd^4\hat\sigma}
                      {\dd\hat t_+ \dd\hat t_- \dd\hat u_+ \dd\hat u_-}
\end{eqnarray}
where $\hat\sigma$ is the Born level cross section of the hard process,
$\Gamma(x_i,Q^2)$ are the structure functions for initial state
radiation, $D(y_i,Q^2)$ are the fragmentation functions used for final
state radiation, and $Q^2$ is the factorization scale.  The invariants
are defined as follows
\begin{eqnarray}
\label{eq:invariants-first}
  s     & = & \left( p_+ + p_- \right)^2 \\
  \hat s  & = & \left( x_+ p_+ + x_- p_- \right)^2 = x_+ x_- s \\
  t_\pm & = & \left( p_\pm - p'_\pm \right)^2 \\
  \hat t_\pm & = & \left( x_\pm p_\pm - \frac{p'_\pm}{y_\pm} \right)^2
         = \frac{x_\pm}{y_\pm} t_\pm \\
  u_\pm & = & \left( p_\mp - p'_\pm \right)^2 \\
  \hat u_\pm & = & \left( x_\mp p_\mp - \frac{p'_\pm}{y_\pm} \right)^2
         = \frac{x_\mp}{y_\pm} u_\pm
\label{eq:invariants-last}
\end{eqnarray}
where the electron mass has been neglected.
In the absence of photon radiation we have of course $t_+ = t_-$ and
$u_+ = u_-$.

Here we restrict ourselves to the leading log approximation (LLA)
where the structure functions $\Gamma(x,Q^2)$ and $D(x,Q^2)$ are
identical.  The structure functions $D$ satisfy the evolution
equation~\cite{Li74,AP77,KF85}
\begin{eqnarray}
\label{eq:DGLAP}
   Q^2 \frac{\partial}{\partial Q^2} D(x,Q^2)
      & = & \frac{\alpha}{2\pi}
             \int\limits_x^1 \frac{\dd z}{z}
                 \left[P_{\mathrm{e}\mathrm{e}}(z)\right]_+
                 D\left(\frac{x}{z},Q^2\right) \\
\label{eq:splitting-function}
  P_{\mathrm{e}\mathrm{e}}(z) & = & \frac{1+z^2}{1-z}
\end{eqnarray}
with initial condition
\begin{equation}
  D(x,m_{\mathrm{e}}^2) = \delta(1-x)
\end{equation}

An explicitly regularized version of (\ref{eq:DGLAP}) which is used in
the Monte Carlo implementation is given by
\begin{eqnarray}
\label{eq:regularized-DGLAP}
   Q^2 \frac{\partial}{\partial Q^2} D(x,Q^2)
     & = & \frac{\alpha}{2\pi}
              \int\limits_x^{1-\epsilon} \frac{\dd z}{z}
                 P_{\mathrm{e}\mathrm{e}}(z)
                 D\left(\frac{x}{z},Q^2 \right) \\
     &   & \mbox{} - \frac{\alpha}{2\pi}
             \left[ \int\limits_0^{1-\epsilon}
              \dd z P_{\mathrm{e}\mathrm{e}}(z) \right]
                  D(x,Q^2). \nonumber
\end{eqnarray}


\subsection{The Algorithms for Multiphoton Radiation}

\subsubsection{Initial State Radiation}

{}From (\ref{eq:invariants-first}--\ref{eq:invariants-last}) it is
obvious that the implementation of (\ref{eq:factorization}) in a Monte
Carlo event generator amounts to solving (\ref{eq:regularized-DGLAP}) by
iteration and taking into account the energy loss in the hard cross
section.

The initial state branching algorithm, which is implemented as a photon
shower algorithm, has already been described in ref.~\cite{ADM+92a}.  It
has been taken over essentially unchanged, the only modification is that
we require conservation of four-momentum at every emission vertex and
allow the electron momentum to take on off-shell values after initial
state radiation.  The off-shell momenta of the electron and positron
after initial state branching are then used as the input momenta for the
subgenerator of the hard Bhabha cross section.  In the present version,
the factorization scale used for initial state radiation is fixed to
$Q^2 = s$.

\subsubsection{Final State Radiation}

In the vicinity of the $\Z$ peak and for not too small scattering
angles, the cross section is dominated by $s$-channel $\Z$ exchange.
Therefore we use the following approach.  If we neglect interference
between initial- and final-state radiation, the radiative corrections in
the final state will then be well approximated by the radiative
corrections to $\Z$ decay.

The differential decay rate for one-photon emission $\Z \to \epem
\gamma$ is related to the decay $\Z \to \epem$ at lowest
order~\cite{Kle86}:
\begin{eqnarray}
  \frac{\dd\Gamma^{(3)}}{\dd x_1 \dd x_2 \dd\Omega_{1,2}}
        & = &
  C_1(x_1,x_2) \cdot \frac{\dd\Gamma^{(2)}}{\dd\Omega_1} +
  C_2(x_1,x_2) \cdot \frac{\dd\Gamma^{(2)}}{\dd\Omega_2}
\label{eq:FSR-3-Jets}
\\
\noalign{\hbox{where}}
  C_1(x_1,x_2) & = & \frac{\alpha}{2\pi} \frac{x_1^2}{(1-x_1)(1-x_2)}
  \nonumber \\
  C_2(x_1,x_2) & = & \frac{\alpha}{2\pi} \frac{x_2^2}{(1-x_1)(1-x_2)}
  \nonumber
\end{eqnarray}
Here $\dd\Gamma^{(2,3)}$ denotes the differential decay width of $\Z$
into $\epem$ ($\epem\gamma$), $\dd\Omega_{1,2}$ refers to the direction
of the outgoing $\mathrm{e}^+$ ($\mathrm{e}^-$) in the $\Z$ rest frame,
and $x_1$ and $x_2$ are the energy fractions of $\mathrm{e}^+$ and
$\mathrm{e}^-$ for the radiative decay.  For a more detailed description
we refer the reader to \cite{Kle86}.

It is easily verified that, in the limit of emission of soft photons or
photons (almost) collinear with the outgoing fermions,
eq.~(\ref{eq:FSR-3-Jets}) reduces to
\begin{equation}
\label{eq:FSR-collinear-limit}
  \frac{1}{\Gamma^{(2)}}
  \frac{\dd\Gamma^{(3)}}{\dd x_1 \dd x_2} \; \sim \; \hat{s}^2
  \frac{x_1^2 + x_2^2}{(2 p_+' \cdot k) (2 p_-' \cdot k)} \; ,
\end{equation}
which exhibits the leading soft and collinear singularities that have to
be resummed to all orders.

We choose the phase space boundary such that the leading logarithms are
correctly reproduced,
\begin{eqnarray}
\label{eq:FSR-phase-space-1}
    1 - x_3(1-\eta) < x_{1,2} < 1 - x_3 \eta
         & , &
        \epsilon < x_3 < 1
\\
\noalign{\hbox{with}}
     x_1 + x_2 + x_3 & = & 2 \; . \nonumber
\end{eqnarray}
The parameter $\eta$ is therefore related to the factorization scale
$Q^2$ associated with final state radiation:
\begin{equation}
\label{eq:FSR-scale-tau}
  \ln \left(\frac{1}{\eta} \right) =
  \ln \left(\frac{Q^2}{m_e^2} \right) - 1
\end{equation}

In the case of multiphoton final state radiation approximations are
necessary.  Our main concern is the proper infrared and collinear limit
of the effective radiation matrix element, plus a good approximation for
the photon that carries the largest transverse momentum relative to the
outgoing fermions, while we want to neglect the interference between
successive emissions, and assume that the radiated photons are widely
separated in phase space.  One can easily see that iteration of the
single photon emission algorithm satisfies these criteria {\em and}
reproduces the leading logarithms, provided the following {\em
virtuality ordering conditions} are met.  Denote by $x_1^{(i)}$,
$x_2^{(i)}$, $i=1,\ldots,n$, the phase space variables for the case of
emission of $n$ photons.  The phase space variables are then required to
satisfy
\begin{equation}
\label{eq:virtuality-condition}
        \min \left[ 1-x_1^{(1)}, 1-x_2^{(1)} \right] >
        \min \left[ 1-x_1^{(2)}, 1-x_2^{(2)} \right] >
                \cdots >
        \min \left[ 1-x_1^{(n)}, 1-x_2^{(n)} \right]
\end{equation}
which are the analogue of the ordering conditions (15) resp.\ (17) of
ref.~\cite{ADM+92a}, except that they are reversed.  This, together with
the phase space (\ref{eq:FSR-phase-space-1}), guarantees the proper
normalization of the leading logarithmic terms.
\begin{eqnarray}
\lefteqn{ \!\!\!\!\!\!\!\!\!\!\!\!\!\!\!\!\!\!\!\!
\int\limits_{\mathrm{PS}} \;
 \prod_{i=1}^n \dd x_1^{(i)} \dd x_2^{(i)}
 \left[ C_1(x_1^{(i)},x_2^{(i)}) + C_2(x_1^{(i)},x_2^{(i)}) \right]
}
\nonumber \\
& = &
 \frac{1}{n!} \left\{ \int \dd x_1 \dd x_2
                     \left[C_1(x_1,x_2) + C_2(x_1,x_2) \right] \right\}^n
\nonumber \\
& = &
 \frac{1}{n!} \left\{ \frac{\alpha}{\pi}
        \left[ \ln\frac{1}{\eta} \left( 2\ln\frac{1}{\epsilon} - \frac{3}{2}
        \right) - \frac{1}{2} \right] \right\}^n
\nonumber \\
& = &
\label{eq:FSR-photon-number}
 \frac{1}{n!} \left( \bar{n}_f \right)^n
\end{eqnarray}

The algorithm for final state multiphoton radiation is thus defined as
follows.  i) Determine the number of photons $\bar{n}_f$ emitted from
the final state according to (\ref{eq:FSR-photon-number}).  ii) In the
case of single-photon emission, use distribution (\ref{eq:FSR-3-Jets}),
with phase space limits (\ref{eq:FSR-phase-space-1}).  iii) For emission
of multiple photons, iterate the algorithm for single photon radiation,
taking into account the phase space conditions
(\ref{eq:virtuality-condition}).


\subsection{The Effective Born Amplitude}
\label{sec:born-amplitudes}

The calculation of the effective Born amplitude follows the BHM/WOH
approach as described in \cite{Hol90,BHP95}.  The amplitude for $\epem
\to \epem$ including virtual corrections can be cast into a form close
to the lowest order amplitude:
\begin{equation}
\label{eq:eff-Born}
  A(\epem \to \epem) =
  A_s^{(\gamma)} + A_s^{(\Z)} + A_t^{(\gamma)} + A_t^{(\Z)} + (box) \; ,
\end{equation}
where $A^{(\gamma)}$ denotes the dressed photon, $A^{(\Z)}$ the dressed
$\Z$ exchange amplitudes, and $(box)$ the terms from box diagrams, which
can be essentially neglected around the $\Z$ peak.

The dressed photon exchange amplitude is written as
\begin{equation}
  A_s^{(\gamma)} =
  \frac{e^2}{1 + \hat{\Pi}^\gamma(s)} \frac{Q_e^2}{s}
  \left[(1 + F_V^{\gamma e}) \gamma_\mu -
             F_A^{\gamma e}  \gamma_\mu\gamma_5\right]
        \otimes
  \left[(1 + F_V^{\gamma e}) \gamma^\mu -
             F_A^{\gamma e}  \gamma^\mu\gamma_5\right] \; ,
\end{equation}
while the $\Z$ exchange amplitude is given by
\begin{eqnarray}
  A_s^{(\Z)} & = &
  e^2 D_\Z(s)
  \left[(v_e + F_V^{\Z e} + Q_e \hat\Pi^{\gamma\Z}(s)) \gamma_\mu -
        (a_e + F_A^{\Z e}) \gamma_\mu\gamma_5\right]
        \otimes
\nonumber \\ && \hspace*{15mm}
  \left[(v_e + F_V^{\Z e} + Q_e \hat\Pi^{\gamma\Z}(s)) \gamma^\mu -
        (a_e + F_A^{\Z e}) \gamma^\mu\gamma_5\right]
\end{eqnarray}
Here $\hat\Pi^{\gamma\Z}(s) = \hat\Sigma^{\gamma\Z}(s) /
(s+\hat\Sigma^{\gamma\gamma}(s))$.  For more details on the notation we
refer the reader to the literature.


\subsection{QED Virtual Corrections}
\label{sec:virtual-corrections}

Besides the weak virtual corrections that are easily included in the
effective Born amplitude (\ref{eq:eff-Born}), we have to include the QED
virtual corrections (vertex corrections, box corrections).  In addition,
we have to properly include the contributions from soft-photon emission
beyond leading logs.

The method of inclusion of nonleading QED corrections in the structure
function formalism is well known (see e.g.\ ref.~\cite{GN90}).  However,
to preserve a fully factorized form of the cross section similar to
(\ref{eq:factorization}), which is mandated by our implementation of
initial- and final-state radiation, we refrain from introducing
$K$-factors and proceed as follows.

The effective Born amplitude is multiplied by (helicity-dependent)
form-factors, as given explicitly e.g.\ in the appendix of
ref.~\cite{BBM91a}.  The photon-mass singularities in the form-factors
are explicitly canceled against the soft-photon contribution.  The
infrared-singular parts are factored out and absorbed into the structure
functions and fragmentation functions, thus uniquely determining an
infrared-finite nonleading correction.

The cross section determined by this matching procedure in principle
fully reproduces the $O(\alpha)$ corrections in the soft limit.
However, since the actual photon shower algorithms do not yet include
initial-final interference, we have to drop the infrared-finite pieces
of the QED boxes and corresponding terms from the soft-photon
corrections for a consistent treatment.


\subsection{Implementation of the Bhabha Cross Section}
\label{sec:born-implementation}

The differential cross section for the hard subprocess is implemented in
a standard way, using importance sampling techniques.  As a majorant
function for the differential cross section, we use
($\tau \equiv -\hat t/\hat s = (1-\cos\theta_{\mathrm{cms}})/2$)
\begin{equation}
  \frac{\dd\bar\sigma}{\dd\tau}(s,\tau) = A(s) + \frac{B(s)}{\tau^2} \;.
\end{equation}
The functions $A(s)$ and $B(s)$ are chosen such that the peaking
behaviour of the full cross section for small angles as well as the
resonant behaviour for large angles in the vicinity of the $\Z$ peak
is efficiently reproduced:
\begin{eqnarray}
  A(s)  & = &
  \frac{4\pi\alpha(s)^2}{s^2} \cdot
  \frac{\chi_\Z^2(s)}{2}
  \left[
  \left(v_{\mathrm{e}}^2 - a_{\mathrm{e}}^2 + \frac{s-M_\Z^2}{s}\right)^2
  + \frac{M_\Z^2 \Gamma_\Z^2}{s^2} \right]
  \\
  B(s)  & = &
  \frac{4\pi\alpha(s)^2}{s^2} \cdot
  \left[ 1 + C \chi_\Z^2(s)
  \left(v_{\mathrm{e}}^2 + a_{\mathrm{e}}^2\right)
  \frac{|s-M_\Z^2|}{s} \right]
  \\
\noalign{\hbox{where}}
  \chi_\Z^2(s)  & = &
  \frac{s^2}{(s-M_\Z^2)^2 + M_\Z^2 \Gamma_\Z^2}
\end{eqnarray}
Here $\alpha(s)$ is the running QED coupling constant, and
$v_{\mathrm{e}}$ and $a_{\mathrm{e}}$ denote the vector and axial vector
couplings of the electron to the $\Z$.  The parameter $C \approx 0.1$
has been empirically adjusted so that $\dd\bar\sigma/\dd\tau$ is a
majorant of the effective Born cross section.

The angular distribution is generated by a combined mapping and
rejection algorithm with mapping function
\begin{equation}
   \frac{A(s)}{B(s)} \, \tau - \frac{1}{\tau} = \chi(\tau)
         = \chi(\tau_{\max}) \cdot \rho +
                 \chi(\tau_{\min}) \cdot (1-\rho) \; ,
\end{equation}
and corresponding event weight,
\begin{equation}
   w = \frac{\dd\sigma(s,\tau)}{\dd\bar\sigma(s,\tau)} \; ,
\end{equation}
where $\dd\sigma$ is calculated using the effective Born amplitude, as
described in the previous sections.


\section{Implementation of \UNIBAB/ \.{\Version/}}
\label{sec:implementation}

Like almost all Monte Carlo event generators, \UNIBAB/ is divided
into three parts: initialization, generation, termination. These are
described in this section.

\subsection{Initialization of \UNIBAB/}

The initializations in \UNIBAB/ are used for computing the value of
variables that will be used frequently during event generation.  The
primary example is the calculation of electroweak couplings, widths, and
masses from a set of input parameters.  To this end, the generic
initialization routine \.{ubinit} calls the subroutine \.{ubigsw} which
contains a simple interface to the initialization routine of the
electroweak library.

Furthermore, internal steering parameters are derived from the cuts
specified by the user.  Finally, a standard \hepevt/ initialization
record is written \cite{AKV89}, which can be read by the analysis
program.

\subsection{Event Generation}

The routine \.{ubgen} produces an event on every call.  The four momenta
are written to a standard \hepevt/ \cite{AKV89} event record, where they
can be read by user supplied analyzers (the default configuration uses
\hepawk/ \cite{Ohl92a}).  See section \ref{sec:f77} for details on
\UNIBAB/'s use of \hepevt/.

After generation of the initial state radiation by the branching routine
\.{ubbini}, a raw Bhabha event corresponding to the approximate cross
section $\bar\sigma$ (\.{ubxtot}) is generated at reduced lepton
energies with the corresponding angular distribution (\.{ubgt}).  This
event will be given a weight according to the ratio of the full
differential cross section including weak corrections (\.{ubxdif}) to
the approximate one (\.{ubxtri}).

Next, the routine \.{ubgppr} constructs the four-vectors of the leptons
after the hard scattering in the effective c.m.s.\ frame.  Finally,
after generation of the final state radiation by the branching routine
\.{ubbfin}, the event will be accepted in \.{ubgacc} if it passes the
given experimental cuts, otherwise it will be rejected and the above
algorithm repeated.

\subsection{Termination}

The cross section for the generated events, and its error, are obtained
from the standard formulae
\begin{eqnarray}
  \sigma_{\mathrm{tot}}(s) & = & \max_{s'>s_0} \{\bar\sigma(s')\}
   \cdot \frac{\mbox{\# of successful trials}}{\mbox{total \# of trials}}
   \\
  \Delta\sigma_{\mathrm{tot}}(s) & = & \max_{s'>s_0} \{\bar\sigma(s')\}
   \cdot \mbox{} \\
 & & \sqrt{\frac{(\mbox{total \# of trials} - \mbox{\# of successful
   trials}) \cdot \mbox{\# of successful trials}}
   {(\mbox{total \# of trials})^3}} \nonumber
\end{eqnarray}
(where $s_0$ is the minimal invariant mass squared of the outgoing
electron positron pair as determined from the given cuts)
and placed into \hepevt/ for inspection by the analysis program, which
might use it to normalize its histograms at this point.


\section{Parameters}
\label{sec:parameters}

The parameters controlling \UNIBAB/ version \Version/ are summarized
in table \ref{tab:unibab-parm}.

\begin{table}
  \begin{minipage}{\textwidth}
  \begin{center}
  \begin{tabular}{|c|c|c|}
    \hline\hline
    Variable name   & semantics             & Default value
    \\\hline\hline
    \.{ahpla}       & $1/\alpha_{\mathrm{QED}}$ & 137.0359895
    \\\hline
    \.{mass1e}      & $m_{\mathrm{e}}$  & $0.51099906\cdot 10^{-3}$ GeV
    \\\hline
    \.{mass1z}      & $M_{\Z}$              & 91.1887 GeV
    \\\hline
    \.{mass1t}      & $m_{\mathrm{t}}$      & 174.0 GeV
    \\\hline
    \.{mass1h}      & $M_{\mathrm{H}}$      & 300.0 GeV
    \\\hline
    \.{alphas}      & $\alpha_S(M_\Z)$      & 0.124
    \\\hline
    \.{ebeam}       & $\mathrm{e}^\pm$ beam energy     & 46 GeV
    \\\hline
    \.{epol}        & $\mathrm{e}^-$ beam polarization & 0
    \\\hline
    \.{ppol}        & $\mathrm{e}^+$ beam polarization & 0
    \\\hline
    \.{ctsmin}      & minimum $\cos \theta^*$          & $-0.9$
    \\\hline
    \.{ctsmax}      & maximum $\cos \theta^*$          & $+0.9$
    \\\hline
    \.{acocut}      & maximum $\epem$ acollinearity angle      & $20^\circ$
    \\\hline
    \.{ecut}        & minimum outgoing $\mathrm{e}^\pm$ energy & 20 GeV
    \\\hline
    \.{evisct}      & minimum invariant mass of final state    & 0
    \\\hline
    \.{nevent}      & Number of events                 & 1000
    \\\hline
    \.{tchann}      & Switch for $t$-channel diagrams  & .true.
    \\\hline
    \.{qedvtx}      & Switch for QED vertex corrections  & .true.
    \\\hline
    \.{qedbox}      & Switch for QED box corrections  & .false.
    \\\hline
    \.{weak}        & Switch for weak corrections   & .true.
    \\\hline
    \.{boxes}       & Switch for weak box diagrams  & .true.
    \\\hline
    \.{isrtyp}      & Key for initial state radiation & 1
    \\\hline
    \.{fsrtyp}      & Key for final state radiation   & 3
    \\\hline
    \.{epsiln}      & Internal infrared cutoff & $10^{-5}$
    \\\hline
    \.{rseed}       & Random number seed       & 54217137
    \\\hline
    \.{errmax}      & maximum error count      & 100
    \\\hline
    \.{verbos}      & verbosity                & 0
    \\\hline
    \.{runid}       & run identification       &
    \\\hline
    \.{stdin}       & standard input           & 5
    \\\hline
    \.{stdout}      & standard output          & 6
    \\\hline
    \.{stderr}      & standard error           & 6
    \\\hline
  \end{tabular}
  \end{center}
  \end{minipage}
  \caption{Parameters controlling \UNIBAB/.  These range from standard
           model parameters ($M_{\Z}, m_{\mathrm{t}}, M_{\mathrm{H}},
           \ldots$), over experimental cuts
           to purely technical parameters like input/output units.}
\label{tab:unibab-parm}
\end{table}

\subsection{Electroweak Parameters}
\label{sec:ew-params}

Starting with \UNIBAB/ version 2.0, the electroweak corrections are
included through calls to an electroweak library.

The electroweak library of \UNIBAB/ \Version/ is based on the library
that is distributed with \ALIBABA/ \cite{BBM91a} and assumes the minimal
standard model.  This library is based on ref.~\cite{Bee89}, which uses
the on-shell renormalization scheme \cite{Hol90}.  The input parameters
are the masses of the $\Z$ (\.{mass1z}), the top-quark (\.{mass1t}) and
the Higgs (\.{mass1h}).  During initialization, the mass of the
$\mathrm{W}$ (\.{mass1w}) as well as $\Gamma_\Z$ (\.{gamm1z}) and
$\sin^2\theta_W$ (\.{sin2w}) are calculated from these parameters.  As
of version 2.2, the electroweak library has been updated to include
QCD-corrections to the $\Gamma_\Z$ and the leading
$m_{\mathrm{t}}^4$-corrections to the $\rho$-parameter (see e.g.,
\cite{BHP95}).

\UNIBAB/ has several switches that control the ``physics'' entering the
hard scattering cross section.  The first, \.{tchann}, may be used to
switch off the contributions of the $t$-channel diagrams to the cross
section, which is useful for the application of the so-called
``$t$-channel subtraction'' to experimental data, and to compare with
programs that describe the process $\epem\to\mu^+\mu^-$ including higher
order QED corrections, like e.~g.\ {\tt KORALZ} \cite{KORALZ}.  With
\.{tchann} set to \.{.false.}, the program effectively simulates the
process $\epem \to \mu^+\mu^-$; however, this mode of operation is not
very efficient.

The switches \.{qedvtx} and \.{qedbox} control the inclusion of the
finite pieces of the QED vertex and box corrections.  The QED vertex
corrections are enabled by default, however, as explained in
section~\ref{sec:virtual-corrections}, the box corrections are disabled.
The box corrections may be switched on to estimate the size of the
neglected initial-final interference effects, but should not be used for
physics simulations, since initial-final interference is not included
for hard photon emission.

The next switch, \.{weak}, may be used to switch off the pure weak
corrections in the electroweak library.  According to standard
terminology, this includes the propagator corrections to the exchanged
photon and $\Z$.  It should be noted, however, that
$\Gamma_{\mathrm{Z}}$ as well as $\sin^2\theta_W$ are computed from the
input parameters even if the weak corrections are switched off.  A final
switch, \.{boxes}, controls the inclusion of the pure weak box diagrams.
The relative contribution of these diagrams is of the order of at most
0.1--0.2\% in the energy range of LEP, whereas their computation may be
quite time consuming.

\subsection{Control of QED corrections}

The photon shower algorithms are controlled by the following switches:
\begin{itemize}
  \item{} \.{isrtyp}: Key for initial state radiation
  \begin{itemize}
    \item[0] Initial state radiation disabled.
    \item[1] Initial state radiation enabled, factorization scale
             $Q^2=s$.
  \end{itemize}
  \item{} \.{fsrtyp}: Key for final state radiation
  \begin{itemize}
    \item[0] Final state radiation disabled.
    \item[1] Final state radiation enabled, factorization scale
             $Q^2=\hat{s}$  (c.f.\ (\ref{eq:FSR-scale-tau})).
    \item[2] Same, but with $Q^2 = -\hat{t} \cdot e$, so that
             $\ln (1/\eta) = \ln(-\hat{t}/m_{\mathrm{e}})$
    \item[3] Same, but with $Q^2 = \hat{s}\hat{t}/\hat{u}$
  \end{itemize}
\end{itemize}
The different choices for the factorization scale for final state
radiation, which are of course completely equivalent at the leading
logarithmic level, can be used to estimate the effect of the missing
initial-final interference effects in \UNIBAB/.

The remaining QED corrections (vertex, boxes) are described in
subsection~\ref{sec:ew-params}.

The former switch \.{bstyle} (\UNIBAB/ versions 2.0 and 2.1) is no
longer avaible, it has been superseeded by \.{isrtyp} and \.{fsrtyp}.

\subsection{Cuts}

The region of phase space where \UNIBAB/ generates events is controlled
by specifying kinematical cuts.  For the sake of simplicity and
efficiency, there are five parameters:
\begin{itemize}
  \item{} \.{ctsmin}: minimum $\cos\theta^*$ (see below),
  \item{} \.{ctsmax}: maximum $\cos\theta^*$,
  \item{} \.{acocut}: maximum acollinearity angle of the outgoing
                      leptons,
  \item{} \.{ecut}:   minimum energy of either of the outgoing leptons,
  \item{} \.{evisct}: minimum invariant mass of final state.
\end{itemize}
The meaning of the angle $\theta^*$ is the following.  Let $\theta_+$
and $\theta_-$ be the angle betweeen the outgoing positron and
electron and the incoming electron beam, respectively.  The angle
$\theta^*$ is then defined by
\begin{equation}
  \cos \theta^* =
        \frac{\cos\left[\frac12(\theta_- + \pi - \theta_+)\right]}
             {\cos\left[\frac12(\theta_- - \pi + \theta_+)\right]}
\end{equation}
If all emitted photons are emitted strictly collinearly with respect to the
radiating lepton, $\theta^*$ equals to the scattering angle in the
partonic subsystem.

Since \UNIBAB/ is designed for the calculation of QED corrections in the
LLA in the large angle regime, the angular cuts are restricted by the
requirement that the factorization scale be the c.m.s.\ energy $s$, so
that terms of the order $\alpha/\pi \cdot \log(-t/s)$ and $\alpha/\pi
\cdot \log(-u/s)$ remain small compared to the leading logarithmic terms
$\alpha/\pi \cdot \log(s/m_{\mathrm{e}}^2)$.  We require (cf.\
eqs.~(\ref{eq:invariants-first}--\ref{eq:invariants-last}))
\begin{equation}
  \left|\alpha/\pi \cdot \log(-\hat t/\hat s)\right| , \;
  \left|\alpha/\pi \cdot \log(-\hat u/\hat s)\right|  < 1\% \; .
\end{equation}
Therefore, the angular cuts in \UNIBAB/ are restricted to the range
\begin{equation}
  -0.985 \leq \cos\theta^*_{\min} < \cos\theta^*_{\max} \leq 0.985 \; ,
\end{equation}
which corresponds to
\begin{equation}
  10^\circ \; < \; \theta^* \; < \; 170^\circ
\end{equation}

While the parameter \.{ecut} has a physical meaning, the cut \.{evisct}
has a technical interpretation and is to be understood in the following
way.  Due to the implementation, each radiated photon may be associated
with either initial state or final state.  Thus one can define an
``invariant mass of the final state''.  It should be noted, however,
that this quantity is unphysical and should be used only for
optimization of the Monte Carlo speed for a given set of physical cuts.
An example is the calorimetric measurement of the Bhabha cross section,
where one does not discriminate between the outgoing lepton and an
accompanying photon if the latter lies within a narrow cone of the
former, but where one cuts on the sum of their energies.

Although most of the above-mentioned cuts have a simple physical
meaning, their implementation in the Monte Carlo generator is not
without problems.  For strictly collinear photon emission, the relations
(\ref{eq:invariants-first}--\ref{eq:invariants-last}) between laboratory
variables and the variables for the hard scattering subprocess allow a
straightforward derivation of the Monte Carlo's internal steering
parameters from these cuts in a strict manner.  If one allows for finite
angles between the photon and the radiating lepton, these relations get
modified, and one has to weaken the internal cuts; in fact, one cannot
even derive these parameters any more in a strict manner, since the
multi-particle phase space becomes too complicated.  In the actual
implementation, we have used empirical relations to derive these
internal steering parameters from those valid for collinear emission, so
that the contribution to the systematic error on inclusive observables
like cross sections is well below the overall precision, which is of the
order of 1\%.  For less inclusive observables, we would recommend that
the user convinces himself by a variation of the cuts that the generated
distribution does not change in the region of interest.

It should be noted that setting the lepton energy cut \.{ecut} to very
small values or setting the acollinearity cut \.{acocut} close to
$180^\circ$ makes the Monte Carlo generator very slow.  The latter also
implies that non-logarithmic terms, which are not implemented in
\UNIBAB/, become more important.

\subsection{Monte Carlo Parameters}

The remaining, more technical Monte Carlo parameters should be almost
self explaining.  Since our photon shower algorithm automatically
includes soft photon exponentiation, the results will not depend on the
value of the internal infrared cutoff \.{epsiln} (which is measured in
units of the beam energy), provided it is kept {\em well below\/} the
experimental energy resolution.  However, it is not advisable to set it
many orders of magnitude lower than the default value, because this may
result in too high photon multiplicities which will overflow internal
tables.


\section{\f77/ Interface}
\label{sec:f77}

\UNIBAB/ version \Version/ provides two application program interfaces
on different levels.  The higher (preferred) level consists of the
command interpreter \.{ubdcmd} which accepts commands in form of
\.{character*(*)} strings.  This driver communicates with the analyzer
\hepawk/ \cite{Ohl92a} by default.  The lower level consists of two
\f77/ subroutine calls: \.{ubpsrv} and \.{unibab}.

\subsection{Higher Level Interface}
\label{sec:f77-high-level}

\begin{example}{\f77/ interface}{ex:f77}
\.{* unibabappl.f}                                                \\
\>\>\> \.{...}                                                    \\
\>\>\> \.{call ubdcmd ('init')}   \C initialize the generator     \\
\>\>\> \.{...}                                                    \\
\>\>\> \.{call ubdcmd ('generate 10000')}
                                   \C generate 10000 events       \\
\>\>\> \.{...}                                                    \\
\>\>\> \.{call ubdcmd ('close')}  \C cleanup                      \\
\>\>\> \.{...}
\end{example}

The simple commands understood by \.{ubdcmd} are  (here keywords are
typeset in typewriter font and variables in italics; vertical bars
denote alternatives)
\begin{itemize}
  \item \.{initialize} \hfil\goodbreak
    Force initialization of \UNIBAB/ and write an initialization
    record into the
    \hepevt/ event record, which should trigger the necessary
    initializations in the analyzer.
  \item \.{generate} $[n]$ \hfil\goodbreak
    Generate \.{nevent} events and call \hepawk/ to analyze them.
    If the optional parameter $n$ is supplied, \.{nevent} is set
    to its value.
  \item \.{close} \hfil\goodbreak
    Write a termination record to \hepevt/, which should
    trigger the necessary cleanups in the analyzer.
  \item \.{statistics} \hfill\goodbreak
    Print performance statistics (this is usually only useful for the
    \UNIBAB/ developers, who are tuning internal parameters).
  \item \.{quit} \hfil\goodbreak
    Terminate \UNIBAB/ without writing a termination
    record.
  \item \.{exit}$\vert$\.{bye} \hfil\goodbreak
    Write a termination record and terminate \UNIBAB/.
  \item \.{set} {\it variable\/}
         {\it ival\/}$\vert${\it rval\/} \hfil\goodbreak
    Set physical or internal parameters.  See the table
    \ref{tab:unibab-parm} for
    a comprehensive listing of all variables.  For example, the
    command \verb+'set ahpla 128.0'+ will set the QED fine structure
    constant to $1/128$.
  \item \.{print} {\it variable\/}$\vert$\.{all} \hfil\goodbreak
    Print the value of physical or internal variables.
    Specifying the special variable \.{all} causes a listing
    of all variables known to \UNIBAB/.
  \item \.{debug}$\vert$\.{nodebug} {\it flag\/} \hfil\goodbreak
    Toggle debugging flags.
  \item \.{testran} \hfil\goodbreak
    Test the portability of the random number generator.  We use
    a generator of the Marsaglia-Zaman variety \cite{MZT90},
    which should give
    identical results on almost all machines.
  \item \.{banner} \hfil\goodbreak
    Print a string identifying this version of \UNIBAB/.
  \item \.{echo} {\it message\/} \hfil\goodbreak
    Print {\it message\/} on standard output.
\end{itemize}

Unique abbreviations of the keywords are accepted, i.e.~\verb+'g 1000'+
generates 1000 events.  The tokens are separated by blanks. Blank
lines and lines starting with a \.{\#} are ignored and may be used for
comments.  For portability, only the first 72 characters of each line
are considered.

On UNIX systems \UNIBAB/ reads default startup files {\tt .unibab}
in the user's home directory and the current directory, if they
exist.

\subsection{Lower Level Interface}
\label{sec:f77-low-level}

\begin{example}{Event generation loop}{ex:f77-ubdcmd}
\.{* ubdcmd.f}                                                    \\
\>\>\> \.{subroutine ubdcmd (cmdlin)}                             \\
\>\>\> \.{character*(*) cmdlin}                                   \\
\>\>\> \.{...}                                                    \\
\>\>\> \.{else if (cmdlin.eq.'init')}                             \\
\>\>\> \>\> \.{call unibab (0)}        \C initialize \UNIBAB/     \\
\>\>\> \>\> \.{call hepawk ('scan')}   \C print
                                            initialization record \\
\>\>\> \.{else if (cmdlin.eq.'generate')}                         \\
\>\>\> \>\> \.{do 10 n = 1, nevent}                               \\
\>\>\> \>\> \>\> \.{call unibab (1)}      \C generate an event    \\
\>\>\> \>\> \>\> \.{call hepawk ('scan')} \C analyze the event    \\
\.{10} \>\>\>\>\> \.{continue}                                    \\
\>\>\> \.{else if (cmdlin.eq.'close')}                            \\
\>\>\> \>\> \.{call unibab (2)}        \C get total cross
                                                    section, etc. \\
\>\>\> \>\> \.{call hepawk ('scan')}   \C finalize analysis       \\
\>\>\> \.{else}                                                   \\
\>\>\> \.{...}                                                    \\
\>\>\> \.{end}
\end{example}

The subroutine \.{unibab(code)} has a single integer parameter.
The parameter \.{code} is interpreted as follows:
\begin{itemize}
  \item{} 0: initialize the generator and write an initialization
      record to \hepevt/.
  \item{} 1: generate an event and store it in \hepevt/.  If \UNIBAB/
    has not been initialized yet, the necessary initializations are
    performed, but no initialization record is written.
  \item{} 2: perform final calculations and write the results
      to \hepevt/.
\end{itemize}
Figure \ref{ex:f77-ubdcmd} displays excerpts from a simplified
version of \.{ubdcmd} which make the correspondences between the
two levels of the \f77/ interface explicit.

\UNIBAB/'s parameters can be accessed on the lower level by the
subroutine \.{ubpsrv (result, action, name, type, ival, rval, dval,
lval)}. The parameter is specified by its (lowercase) name in the
\.{character*(*)} string \.{name}.  The string \.{action} is either
\verb+'read'+ or \verb+'write'+ corresponding to whether the parameter
is to be inspected or modified. The type of the parameter
(\verb+'int'+, \verb+'real'+, \verb+'dble'+, or \verb+'lgcl'+) is
returned in \.{type}, if \.{action} is set to \verb+'read'+; if
\verb+action+$=$\verb+'write'+, the type must be specified in \.{type}.
Depending on this type the value is passed in
\.{ival}, \.{rval}, \.{dval}, or \.{lval}, respectively.  The
following error codes will be returned in the string \.{result}:
\verb+' '+: no error, \verb+'enoarg'+: invalid \.{action},
\verb+'enoent'+: no such parameter, \verb+'enoperm'+: permission
denied, and \verb+'enotype'+: invalid \.{type}.

The protection scheme implemented with this parameter handling has
been described in \cite{ADM+92a}.  Its main purpose is to guarantee
consistency of user defined and computed parameters in the generation
phase of the Monte Carlo.

\subsection{Additional information in \hepevt/}

Because \UNIBAB/ uses since version \Version/ the standard \hepevt/
event record internally,
not only stable particles with \.{isthep(i) = 1} will be present.
Adapting the conventions of the \HERWIG/ Monte Carlo \cite{MWA+92},
we use the following status codes
\begin{itemize}
  \item{} 101: $\mathrm{e}^-$ beam,
  \item{} 102: $\mathrm{e}^+$ beam,
  \item{} 103: $\epem$ center of mass system,
  \item{} 110: $\epem$ hard scattering center of mass system,
  \item{} 111: $\mathrm{e}^-$ before hard scattering,
  \item{} 112: $\mathrm{e}^+$ before hard scattering,
  \item{} 113: $\mathrm{e}^-$ after hard scattering,
  \item{} 114: $\mathrm{e}^+$ after hard scattering.
\end{itemize}
However, these entries have {\em no\/} physical significance and
should {\em never\/} be used in any analysis. An exception to this
rule are the beam particles 101 and 102, which are convenient for
defining the reference frame (positive $z$ axis pointing in the direction
of the incoming electron) and are used e.g.~by \hepawk/
\cite{Ohl92a} for this purpose.  Only the particles with status code 1
belong to the final state as predicted by \UNIBAB/.


\section{Conclusions}
\label{sec:concl}

We have presented the new version \Version/ of the Monte Carlo event
generator \UNIBAB/ for Bhabha scattering at LEP/SLC.  The distinguishing
feature of \UNIBAB/ is the inclusion of higher order electromagnetic
corrections, including exponentiated soft photons, in a photon shower
approach.  In contrast to fixed order calculations which have to be
exponentiated by hand, \UNIBAB/ handles the multiphoton effects
explicitly.


\section*{Acknowledgements}

One of the authors (H.A.) would like to thank Guido Montagna and Oreste
Nicrosini for stimulating discussions during the LEP2 workshop.

\appendix

\section{Availability}

The latest release of \UNIBAB/ is available from the authors upon
request.  In addition, \UNIBAB/ is available by anonymous ftp from the
directory\\
{\tt ftp://crunch.ikp.physik.th-darmstadt.de/pub/anlauf/unibab} \\
It is nevertheless recommended to notify the authors, in order to be
informed of future bug fixes and enhancements.

\UNIBAB/ is distributed in PATCHY/CMZ card format \cite{KZ88,CMZ}, but
machine specific and plain \f77/ versions are made available on request.
A more modern (auto-configuring) and self-contained version of the Monte
Carlo generator will be made available in a future release.


\section{Revision History}
\label{sec:History}

\subsection*{Version 2.2, December 1995}

\begin{itemize}
  \item{} Update of the electroweak library.
  \item{} Improved final state photon shower algorithm.
  \item{} Different choices of factorization scale for final state
          radiation.
  \item{} QED virtual corrections.
  \item{} Speed-up of electroweak corrections.
  \item{} Fixed minor bug in initial-state photon shower.
\end{itemize}

\subsection*{Version 2.1, September 1994}

\begin{itemize}
  \item{} Beam polarization.
\end{itemize}

\subsection*{Version 2.0, June 1993}

\begin{itemize}
  \item{} weak corrections included.
  \item{} problems with energy non-conservation of old final-state
        radiation algorithm fixed.
  \item{} more user-friendly: physical cuts replace ``bare'' Monte
        Carlo steering parameters.
\end{itemize}

\subsection*{Version 1.3, September 1992}

\begin{itemize}
  \item{} improved form of initial state radiation, as implemented in
        \cite{ADM+92a}.
  \item{} final state radiation.
\end{itemize}

\subsection*{Version 1.2, July 1992}

Complete internal rewrite
\begin{itemize}
  \item{} \hepevt/ support (also used internally now),
  \item{} double precision arithmetic (which is the native floating
    point mode on most modern machines),
  \item{} new driver program (using \KRONOS/\cite{ADM+92a} routines).
\end{itemize}
The physics algorithms have not changed from version 1.1.  The purpose
of this release is to provide a usable baseline for further
improvements.

\subsection*{Version 1.1, September 1989}

Maintenance release.

\subsection*{Version 1.0, July 1989}

First official release.


\section{Common Blocks and Subroutines}
\label{sec:ext-names}

To avoid possible name clashes with other packages, all external symbols
exported by \UNIBAB/ proper begin with the two letters \.{UB}, except
for the routine \.{unibab} itself and the \hepevt/ common block.  The
names of the subroutines and common blocks of the provided electroweak
library have been taken over unchanged.

\begin{itemize}
  \item Common Blocks: \hfil\goodbreak
    \begin{itemize}
      \item \.{/ubpcom/}: Main parameter common block, holds all physical
        parameters. Application programs should access this common block
        through the \.{ubpsrv} routine
        (cf.~section~\ref{sec:f77-low-level}).
      \item \.{/ubcbrn/}: holds the maximum Born cross section in the
        available energy interval.
      \item \.{/ubcevt/}: passes the particle momenta between
        subroutines.
      \item \.{/ubcmsc/}: internal parameters.
      \item \.{/ubcsta/}: holds statistics on \UNIBAB/'s performance.
      \item \.{/ubctri/}: lookup table for driver keywords.
    \end{itemize}
  \item Primary entry point: \hfil\goodbreak
    \begin{itemize}
      \item \.{unibab}: main entry point, described in
        section~\ref{sec:f77-low-level}.
    \end{itemize}
  \item Branching:
    \begin{itemize}
      \item \.{ubbini}: initial state branching for one lepton.
      \item \.{ubbfin}: final state radiation for one lepton.
    \end{itemize}
  \item Cross sections:
    \begin{itemize}
      \item \.{ubxdif}: differential cross section
        $\dd\sigma/\dd\cos\theta$ (in pbarn).  This function contains
        a very simple interface to the electroweak library.
      \item \.{ubxtri}: approximate differential cross section
        used for mapping.
      \item \.{ubxtot}: integral of the approximate cross section
        used for mapping.
      \item \.{ubxcof}: coefficients of the approximate cross section.
      \item \.{ubffac}: QED vertex form factors and box corrections.
    \end{itemize}
  \item Initialization:
    \begin{itemize}
      \item \.{ubinit}: initialize \UNIBAB/.
      \item \.{ubigsw}: setup standard model parameters.  This
        subroutine also calls the initialization routines of the
        electroweak library and transfers parameters between these two
        modules.
      \item \.{ubibmx}: calculate the maximal Born cross section in
        the considered energy interval.
      \item \.{ubibn}: negative Born cross section.
      \item \.{ubeeni}: enters an initialization record into \hepevt/.
    \end{itemize}
  \item Event generation:
    \begin{itemize}
      \item \.{ubgen}: main event generation routine.
      \item \.{ubgppr}: construct four momenta of hard subprocess in
        the effective c.m.\ system.
      \item \.{ubgt}: generate momentum transfer in the hard
        subprocess.
      \item \.{ubgacc}: checks if current raw event satisfies physical
        cuts.
      \item \.{ubeent}: enter a particle in \hepevt/.
      \item \.{ubenew}: start a new \hepevt/ record.
      \item \.{ubenul}: zero a \hepevt/ entry.
    \end{itemize}
  \item Termination:
    \begin{itemize}
      \item \.{ubclos}: write a final record to \hepevt/.
      \item \.{ubeens}: enters a termination record into \hepevt/.
      \item \.{ubstat}: print statistics on \UNIBAB/'s performance.
    \end{itemize}
  \item Driver program:
    \begin{itemize}
      \item \.{ubdriv}: main driver program, calling the command loop.
      \item \.{ubdloo}: command loop, reading commands from the
        terminal or from files.
      \item \.{ubdcmd}: execute a single command
        (cf.~section~\ref{sec:f77-high-level}).
      \item \.{ubdlxs}: get the next string from the command line.
      \item \.{ubdlxd}: get the next floating point number from the
        command line.
      \item \.{ubdlxi}: get the next integer from the command line.
    \end{itemize}
  \item Parameter management:
    \begin{itemize}
      \item \.{ubpsrv}: inspect or modify parameters
        (cf.~section~\ref{sec:f77-low-level}).
      \item \.{block data ubpini}: initialize parameters with their
        default values.
      \item \.{ubpprn}: print parameters.
    \end{itemize}
  \item General utility routines:
    \begin{itemize}
      \item \.{ubrgen}: random number generator of the
         Marsaglia-Zaman type \cite{MZT90}.
      \item \.{ubrtst}: selftest routine for \.{ubrgen}.
      \item \.{ububoo}: boost a four-vector.
      \item \.{ubulwr}: convert string to lower case.
      \item \.{ubumin}: minimize function.
      \item \.{ubumsg}: print a message.
      \item \.{ubuort}: construct three orthogonal three-vectors for a
        given $z$ direction.
      \item \.{ubupro}: project a pair of four-vectors onto mass shell.
      \item \.{ubutim}: check remaining CPU time.
    \end{itemize}
  \item Keyword search
    (using the dynamic tries described in \cite{Dun91}):
    \begin{itemize}
      \item \.{ubtins}: insert a new keyword.
      \item \.{ubtlup}: look up a (possibly abbreviated) keyword.
      \item \.{ubtnew}: insert a new node into the trie.
      \item \.{ubtlen}: calculate length of keyword.
      \item \.{ubtc2a}: convert keyword from \.{character*(*)} to
        \.{integer(*)}.
    \end{itemize}
  \item Electroweak library entry points:
    \begin{itemize}
      \item \.{ewinit}: initialize electroweak library.
      \item \.{eeeew}: effective cross section including virtual
        corrections.
    \end{itemize}
\end{itemize}



\newpage

\section*{Test Run}

\UNIBAB/ version \Version/ is distributed together with a sample command
file and \hepawk/ script, which are given below.  To run this example,
the user will need to link \UNIBAB/ with the CERN library, because
histogramming is done by HBOOK \cite{HBOOK}.

The file {\tt sample.unibab} is read from standard input ({\tt unit
  stdin}, which is initialized to 5), and {\tt sample.hepawk} is read
from the file {\tt SCRIPT} (i.e.\ under MVS from the file which has been
allocated to the {\tt DDNAME SCRIPT} and under UNIX from the file {\tt
  script} or from the value of the environment variable {\tt SCRIPT}).

After \UNIBAB/ has been successfully built and linked with \hepawk/ on a
supported UNIX system, running either {\tt make check} or {\tt make
test} should validate the build.

\subsection*{{\tt sample.unibab}}

Here is a simple \UNIBAB/ command file, setting up parameters and
generating 10000 events.

{\small
\begin{verbatim}
# sample.unibab -- sample UNIBAB command file

# kinematical cuts (e+e-)
set ctsmin -0.8
set ctsmax  0.8
set ecut     20
set acocut   30
set ebeam    48

# run
init
gen 10000
close
quit
\end{verbatim}
} 


\newpage

\subsection*{{\tt sample.hepawk}}

This is a small \hepawk/ script that plots the acollinearity
distribution for Bhabha events, for which the acollinearity is larger
than 0.01 radians.

{\small
\begin{verbatim}
# sample.hepawk -- sample HEPAWK analyzer for UNIBAB.

BEGIN
  {
    printf ("\nWelcome to the UNIBAB test:\n");
    printf ("***************************\n\n");
    printf ("Monte Carlo Version: %s\n", REV);
    printf ("                Run: %d\n", RUN);
    printf ("               Date: %s\n\n", DATE);

    lower_angle_cut = 20/DEG;
    upper_angle_cut = PI - lower_angle_cut;
    minimum_acollinearity_cut = 0.01;

    h_acollinearity
      = book1 (0, "electron acollinearity", 50, 0, PI/6);

    incut = 0;               # initialize counter
  }

  {
    # Collect the outgoing electron and positron
    $electron = $positron = $NULL;
    for (@l in LEPTONS)
      if (@l:id == _pdg_electron)
        $electron = @l:p;
      else if (@l:id == - _pdg_electron)
        $positron = @l:p;
  }

lower_angle_cut <= angle ($electron, @B1:p) <= upper_angle_cut
 && lower_angle_cut <= angle ($positron, @B1:p) <= upper_angle_cut
  {
    # sample the acollinearity, but cut out Born-like events
    acollinearity = PI - angle ($electron, $positron);
    if (acollinearity > minimum_acollinearity_cut)
      {
        incut++;
        fill (h_acollinearity, acollinearity);
      }
  }

END
  {
    # Dump some numbers
    printf ("\nRESULTS:\n");
    printf ("********\n\n");
    printf ("generated events:   %d, generated cross section: %g nb\n",
            NEVENT, XSECT * 1e6);
    printf ("events within cuts: %d, cross section:           %g nb\n",
            incut, (incut/NEVENT) * XSECT * 1e6);

    printf ("\nHISTOGRAMS:\n");
    printf ("***********\n\n");
    scale (XSECT/NEVENT*1e6);
    plot ();   # plot the histograms

    printf ("\ndone.\n");
  }
\end{verbatim}
} 

\newpage

\subsection*{{\tt sample.output}}

The following output should result from the input files above, modulo
small roundoff errors.

{\small
\begin{verbatim}
ubdcmd: message: Starting UNIBAB, Version 2.01/08, (build 951206/1241)
hepawk: message: starting HEPAWK, Version 1.6

Welcome to the UNIBAB test:
***************************

Monte Carlo Version: v02.01 (Dez 06 00:00:00  1995)
                Run: 1035996352
               Date: Dez 06 13:07:00  1995


RESULTS:
********

generated events:        10000, generated cross section:  .2484     nb
events within cuts:       5904, cross section:            .1467     nb

HISTOGRAMS:
***********

1electron acollinearity

 HBOOK     ID =         1                                        DATE  06/12/95

        1.36     I
        1.32     I      I
        1.28     0     I0
        1.24     I     II
        1.2            0I
        1.16           I
        1.12             I
        1.08         II  0
        1.04         00  I
        1            II
         .96         I
         .92      I I
         .88      II0
         .84      00I
         .8       II      I
         .76              0
         .72              I
         .68
         .64               I
         .6                0
         .56               I
         .52                I
         .48                0
         .44                I
         .4                  I
         .36                 0
         .32                  0
         .28                  I0
         .24                   I
         .2                     I0I0
         .16                    0I0I00 0 I
         .12    0                    I0I000  0I
         .08    I                     I   I00 000000000000 I000II0
         .04                                   II     IIII00II 00I

 CHANNELS  10   0        1         2         3         4         5
            1   12345678901234567890123456789012345678901234567890

 CONTENTS   1.   1   11111
 *10**  2   0   02888001207543221111110101000000000000000000000000
            0   86304019453955855746329290866974455566444423445334
            0   97779369787190039861961166622697994727229927292474
            0   40246609138263740965179798911952767111227743272837

 LOW-EDGE   1.            1111111111222222222333333333344444444455
 *10**  1   0   01234567890123456789013456789012345678901235678901
            0   00011223344556677888990011223344556667788990011223
            0   04948383727161605059493838272616150594948382727161
            0   07429630852964185207418630742963085296418529741853

 * ENTRIES =      11808      * ALL CHANNELS =  .1467E+00      * UNDERFLOW =
.00
 * BIN WID =  .1047E-01      * MEAN VALUE   =  .1140E+00      * R . M . S =
.10

done.
ubdriv: message: bye.
\end{verbatim}
} 



\newpage

\begin{thebibliography}{99}

\bibitem{AKV89}
G.~Altarelli, R.~Kleiss, C.~Verzegnassi, editors,
\newblock {\em {Z} Physics at {LEP} 1}, CERN 89-08, Geneva 1989.

\bibitem{BHP95}
D. Bardin, W. Hollik, G. Passarino, editors,
\newblock {\em Reports of the Working Group on Precision Calculations
for the $\Z$ Resonance}, CERN 95-03, Geneva 1995.

\bibitem{Ols95}
A. Olshevski, {\em EPS Conference on High Energy Physics},
Brussels, July 1995.


\bibitem{BBM91b}
W.~Beenakker, F.A.~Berends, S.C.~van der Marck,
{\em Nucl.~Phys.~\bf B355} (1991) 281.

\bibitem{JRWW92}
S.~Jadach, E.~Richter-W\c{a}s, B.F.L.~Ward, Z.~W\c{a}s,
{\em Comp.\ Phys.\ Comm.} {\bf 70} (1992) 305.

\bibitem{LEP2:BhaWG}
S. Jadach, O. Nicrosini, convenors,
{\em Report of the Subgroup on ``Event Generators for Bhabha Scattering''},
CERN Workshop on Physics at LEP2,
CERN Yellow Report, to appear in 1996.



\bibitem{BBM91a}
W.~Beenakker, F.A.~Berends, S.C.~van der Marck,
{\em Nucl.~Phys.~\bf B349} (1991) 323.

\bibitem{MM92}
M. Martinez, R. Miquel, {\em Z. Phys.} {\bf C53} (1992) 115; \\
P. Comas, R. Martinez, {\em Z. Phys.} {\bf C58} (1993) 15.

\bibitem{zfitter:1992}
D. Bardin et al., FORTRAN program ZFITTER, CERN-TH. 6443/92,
hep-ph/9412201;
{\em Z. Phys.} {\bf C44} (1989) 493,
{\em Nucl.\ Phys.} {\bf B351} (1991) 1,
{\em Phys.\ Lett.} {\bf B255} (1991) 290.

\bibitem{MPN+93}
G.~Montagna, F.~Piccinini, O.~Nicrosini, G.~Passarino, R.~Pittau,
{\em Comp.\ Phys.\ Comm.} {\bf 76} (1993) 328; \\
G.~Montagna, O.~Nicrosini, G.~Passarino, F.~Piccinini, R.~Pittau,
{\em Nucl.\ Phys.} {\bf B401} (1993) 3; \\
G.~Montagna, O.~Nicrosini, G.~Passarino, F.~Piccinini,
CERN-TH.7463/94, to appear in Comp.\ Phys.\ Comm.


\bibitem{sabspv:1995}
M.~Cacciari, G.~Montagna, O.~Nicrosini, F.~Piccinini,
{\em Comp.\ Phys.\ Comm.} {\bf 90} (1995) 301.

\bibitem{bhagen95}
M. Caffo, H. Czyz, E. Remiddi,
{\em Nuovo Cim.} {\bf 105A} (1992) 277,
{\em Int.\ J. Mod.\ Phys.} {\bf 4} (1993) 591,
{\em Phys.\ Lett.} {\bf B327} (1994) 369,
and program BHAGEN95, in preparation.


\bibitem{UNIBAB2.0}
H.~Anlauf, H.D. Dahmen, P.~Manakos, T.~Mannel, H. Meinhard, T.~Ohl,
{\em Comp.\ Phys.\ Comm.} {\bf 79} (1994) 466.

\bibitem{bhagene3:1995}
J.H. Field, Phys.\ Lett.\ B323 (1994) 432; \\
J.H. Field, T.~Riemann, ``BHAGENE3 a Monte Carlo Event Generator
for Lepton Pair Production and Wide Angle Bhabha Scattering in
$\epem$ Collisions near the $\Z$ Peak'',
UGVA-DPNC 1995/6-166, DESY 95-100, to be published in Comp.\ Phys.\ Comm.

\bibitem{bhwide:1995} S. Jadach, W. Placzek, B.F.L. Ward,
``BHWIDE 1.00: $O(\alpha)$ YFS exponentiated Monte Carlo for
Bhabha scattering at wide angles for LEP/SLC and LEP2'',
University of Tennesee preprint UTHEP-95-1001 (unpublished).


\bibitem{DMMO89}
H.D.~Dahmen, P.~Manakos, T.~Mannel, T.~Ohl,
Preprint IKDA 89/28, SI 89-8, June 1989 (unpublished).

\bibitem{DMMO91}
H.D.~Dahmen, P.~Manakos, T.~Mannel, T.~Ohl,
{\em Z.~Phys.} {\bf C50} (1991) 75.

\bibitem{ADM+92a}
H.~Anlauf, H.~D. Dahmen, P.~Manakos, T.~Mannel, T.~Ohl,
{\it Comp.\ Phys.\ Comm.~\bf 70} (1992) 97.

\bibitem{Ohl92b}
T.~Ohl, DESY 92-097, July 1992 (unpublished).


\bibitem{KF85}
E.A.~Kuraev, V.S.~Fadin,
{\em Yad.~Fiz.~\bf 41} (1985) 733.

\bibitem{AM86}
G.~Altarelli, G.~Martinelli, in
J.~Ellis, R.~Peccei (eds.), {\em Physics at LEP\/},
CERN 86-02, Geneva 1986.

\bibitem{BBN89}
W.~Beenakker, F.A.~Berends, W.L.~van Neerven,
Contribution to the Ringberg workshop, April 1989.

\bibitem{Li74}
L.N.~Lipatov, {\em Yad.~Fiz.~\bf 20} (1974) 181.

\bibitem{AP77}
G.~Altarelli, G.~Parisi,
{\em Nucl.~Phys.~\bf B126}, (1977) 298.


\bibitem{Kle86}
R. Kleiss, {\em Phys.\ Lett.} {\bf B180} (1986) 400.

\bibitem{Ohl92a}
T.~Ohl, {\it Comp.~Phys.~Comm.~\bf 70} (1992) 120, and
IKDA 95/15, hep-ex/9504007, April 1995 (unpublished).

\bibitem{Bee89}
W.~Beenakker, Ph.D. thesis, University of Leiden, 1989.

\bibitem{Hol90}
W.~Hollik, {\em Fortsch.~Phys.~\bf 38} (1990) 165.

\bibitem{GN90}
M. Greco, O. Nicrosini, {\em Phys.\ Lett.} {\bf B240} (1990) 219.

\bibitem{KORALZ}
S.~Jadach, B.~F.~L.~Ward, Z.~W\c{a}s,
{\em Comp.~Phys.~Comm.~\bf 66} (1991) 276.

\bibitem{MZT90}
G.~Marsaglia, A.~Zaman, W.-W.~Tsang, {\em Stat.~Prob.~Lett.~\bf 9}
(1990) 35.

\bibitem{MWA+92}
G.~Marchesini, B.~R.~Webber, et.~al.,
{\em Comp.~Phys.~Comm.} {\bf 67} (1992) 465.

\bibitem{Kle90}
R. Kleiss, {\em Nucl.\ Phys.\/} {\bf B347} (1990) 67.

\bibitem{KZ88}
H.~J.~Klein, J.~Zoll,
\newblock {\em {PATCHY Version 4.13}, Reference Guide},
\newblock CERN, Geneva 1988.

\bibitem{CMZ}
{\em CMZ Version 1.41 User's Guide \& Reference Manual}, CodeME S.A.R.L.,
St Genis-Pouilly (France), 1993.

\bibitem{Dun91}
J.~A.~Dundas III, {\em Software--Practice and Experience, \bf 21}
(1991) 1027.

\bibitem{HBOOK}
R. Brun, D. Lienhard,
HBOOK Version 4, User Guide, CERN, Geneva 1987.

\end{thebibliography}
\end{document}